\documentclass[oneside]{IEEEtran}
\usepackage[cmex10]{amsmath}
\usepackage{amssymb}
\usepackage{cite}
\usepackage{amsmath}
\usepackage{amssymb}
\usepackage{booktabs}
\usepackage{balance}
\usepackage{subfigure}
\usepackage{multicol}
\usepackage{graphicx}
\usepackage{epsfig}
\usepackage{epstopdf}
\usepackage{algorithm}
\usepackage[colorlinks,linkcolor=red]{hyperref}
\usepackage{algorithmic}
\usepackage{bm}
\usepackage{amsmath, amssymb, amsthm}
\usepackage{array}
\usepackage{makecell} \usepackage{slashbox}
\usepackage{multirow}
\usepackage{float}
\usepackage{multicol,multicap,graphicx}
\usepackage{1-in-2}
\usepackage{mathrsfs}
\usepackage{url}
\usepackage{ifpdf}
\usepackage[bf]{caption2}
\usepackage{color}

\renewcommand\tablename{Table}
\renewcommand\thetable{\arabic{table}}

\renewcommand{\figurename}{Figure}

\ifCLASSINFOpdf
\else
\fi

\begin{document}
\title{\sc{When xURLLC Meets NOMA: A Stochastic Network Calculus Perspective}}
\author{Yuang~Chen,
        Hancheng~Lu,
        Langtian~Qin,
        Yansha~Deng,
        and Arumugam Nallanathan
}
\maketitle

\section*{Abstract}
The advent of next-generation ultra-reliable and low-latency communications (xURLLC) presents stringent and unprecedented requirements for key performance indicators (KPIs). As a disruptive technology, non-orthogonal multiple access (NOMA) harbors the potential to fulfill these stringent KPIs essential for xURLLC. However, the immaturity of research on the tail distributions of these KPIs significantly impedes the application of NOMA to xURLLC. Stochastic network calculus (SNC), as a potent methodology, is leveraged to provide dependable theoretical insights into tail distribution analysis and statistical QoS provisioning (SQP). In this article, we develop a NOMA-assisted uplink xURLLC network architecture that incorporates an SNC-based SQP theoretical framework (SNC-SQP) to support tail distribution analysis in terms of delay, age-of-information (AoI), and reliability. Based on SNC-SQP, an SQP-driven power optimization problem is proposed to minimize transmit power while guaranteeing xURLLC's KPIs on delay, AoI, reliability, and power consumption. Extensive simulations validate our proposed theoretical framework and demonstrate that the proposed power allocation scheme significantly reduces uplink transmit power and outperforms conventional schemes in terms of SQP performance.

\IEEEpeerreviewmaketitle
\mathfootnote{\centering\footnotesize{Yuang~Chen, Hancheng~Lu, and Langtian~Qin are with University of Science and Technology of China; Yansha~Deng is with King's College London; Arumugam~Nallanathan is with Queen Mary University of London.}}
\section{Introduction}
\par Ultra-reliable and low-latency communication (URLLC) is a newly emerging service in fifth-generation communication systems (5G), pursuant to the 3rd Generation Partnership Project (3GPP) standards, which mandates to meet $99.999\%$ (5-nines) reliability within 1 ms latency target \cite{bennis2018ultrareliable,she2021tutorial}. Nevertheless, the emergence of novel mission-critical and high-risk control applications with more stringent latency, reliability, and freshness, such as robot control, vehicle-to-vehicle communications (V2V), remote surgery, and metaverse, has put forward even more arduous demands on the key performance indicators (KPIs) of URLLC. In this perspective, the sixth-generation communication systems (6G) have brought forth the next-generation URLLC (xURLLC), which necessitates even higher levels of reliability ($99.9999\%$), lower latency (sub-1-ms), and new key performance indicators (KPIs), particularly age-of-information (AoI) and energy efficiency (EE)\cite{park2022extreme,yates2021age,zhang2021statistical}. Achieving these rigorous KPIs requires understanding the statistical characteristics of rare and extreme events, typically also referred to as \emph{tail probability}. Therefore, it appears more meaningful to consider tail probabilities in xURLLC, which is dissimilar to the previous average quality-based network design methodologies\cite{park2022extreme,she2021tutorial,bennis2018ultrareliable}.

\par Non-orthogonal multiple access (NOMA) can be a promising approach for enhancing the service capability of xURLLC's stringent KPIs, as it has emerged as a disruptive technology to significantly improve spectrum efficiency and radio resource utilization\cite{dai2015non,schiessl2020noma}. Various attempts have been made to exploit NOMA to improve the performance of 5G URLLC\cite{liu2022optimization,kotaba2020urllc,yin2022power,schiessl2020noma}. For instance, grant-free NOMA has been introduced to mitigate serious transmission delay and network congestion in massive URLLC scenarios \cite{liu2022optimization}. To achieve ultra-reliability while guaranteeing low latency, NOMA is integrated into hybrid automatic repeat request (HARQ) to actualize highly efficient transmission and retransmission of short packets by allowing concurrent utilization of wireless resources in a non-orthogonal manner \cite{kotaba2020urllc}. Additionally, the high power consumption caused by stringent QoS requirements of URLLC can be effectively improved by using cooperative NOMA for packet re-management by incorporating packet splitting and packet combining\cite{yin2022power}.

\par The aforementioned studies only emphasize average qualities, such as average delay, average throughout, and average packet-loss rate. However, these average quality-based network design methodologies deviate from the design principles mandated by xURLLC, which aim to analyze the tail probabilities of its rigorous KPIs. Unfortunately, existing research on xURLLC has not provided a priori guidance on how to analyze the tail probability of its KPIs, such as delay, age-of-information (AoI), and reliability\cite{park2022extreme,yates2021age,zhang2021statistical}. Additionally, current research on NOMA and xURLLC are mutually parallel\cite{park2022extreme,she2021tutorial,bennis2018ultrareliable,liu2022optimization,kotaba2020urllc,yin2022power}, with the solutions developed for NOMA and 5G-URLLC not necessarily offering direct interoperability for xURLLC research. Thus, challenges persist in achieving a consistent NOMA-assisted transmission scheme for xURLLC that guarantees ultra-reliability, low-latency, and high-freshness (i.e., AoI). Last but not least, the co-channel interference existing in NOMA and the rigorous KPIs for xURLLC make it particularly difficult to design a NOMA-assisted xURLLC network architecture\cite{dai2015non,park2022extreme}.

\par To effectively overcome the aforementioned challenges, it is imperative to develop a theoretical framework that accurately captures the tail distributions of xURLLC's KPIs. Stochastic network calculus (SNC) is a methodology that has the potential to provide dependable theoretical insights into tail distribution analysis and statistical QoS provisioning (SQP) for xURLLC\cite{fidler2014guide,jiang2008stochastic}. The theoretical basis of SNC is $\left(\mathrm{min},\times\right)$-algebra\cite{fidler2014guide}, which enables the analysis of non-asymptotic statistical probabilities of complex stochastic processes, commonly referred to as tail probabilities\cite{fidler2014guide,jiang2008stochastic}. These probabilities are typically formulated as $\mathbb{P}\left[\mathrm{metric} > \mathrm{budget}\right] \leq \varepsilon_{th}$, where the variable $\varepsilon_{th}$ denotes the extremely small violation probability threshold of the sensitive \texttt{metric}. However, leveraging SNC theory to analyze the tail distributions for the integration of NOMA and xURLLC is non-trivial since co-channel interference, statistical QoS requirements, and power allocation need to be comprehensively investigated.

\par In this article, we have proposed an innovative NOMA-assisted uplink xURLLC network architecture (NOMA-xURLLC) that integrates NOMA to significantly enhance the service capability of xURLLC. To overcome the challenges in capturing the tail distributions of xURLLC’s KPIs, we have designed an SNC-based SQP theoretical framework (SNC-SQP), which can provide dependable theoretical insights for the xURLLC’s SQP. In particular, we have defined the statistical delay violation probability (SDVP) and statistical AoI violation probability (SAVP) to characterize the tail distributions of delay and AoI, respectively, and have derived the closed-form expressions of their upper bounds, i.e., UB-SDVP and UB-SAVP. Based on SNC-SQP, we have formulated an SQP-driven power optimization problem that minimizes the uplink transmit power of xURLLC while guaranteeing the statistical QoS requirements in terms of delay, AoI, reliability, and power consumption. Extensive simulations have demonstrated the accurate capturing ability of the designed theoretical framework and have validated that the proposed power allocation scheme outperforms conventional schemes in orthogonal multiple access (OMA) by significantly reducing uplink transmit power and improving SQP performance.

\par The remainder of this article is structured as follows. The system architecture and key challenges of NOMA-xURLLC are first described. Then, we introduce the SNC-SQP theoretical framework. Next, the SQP-driven power optimization problem is formulated and addressed, followed by the performance evaluation. Finally, we conclude this article and discuss future issues.

\begin{figure*}[htbp]
\centering
\includegraphics[scale=0.13]{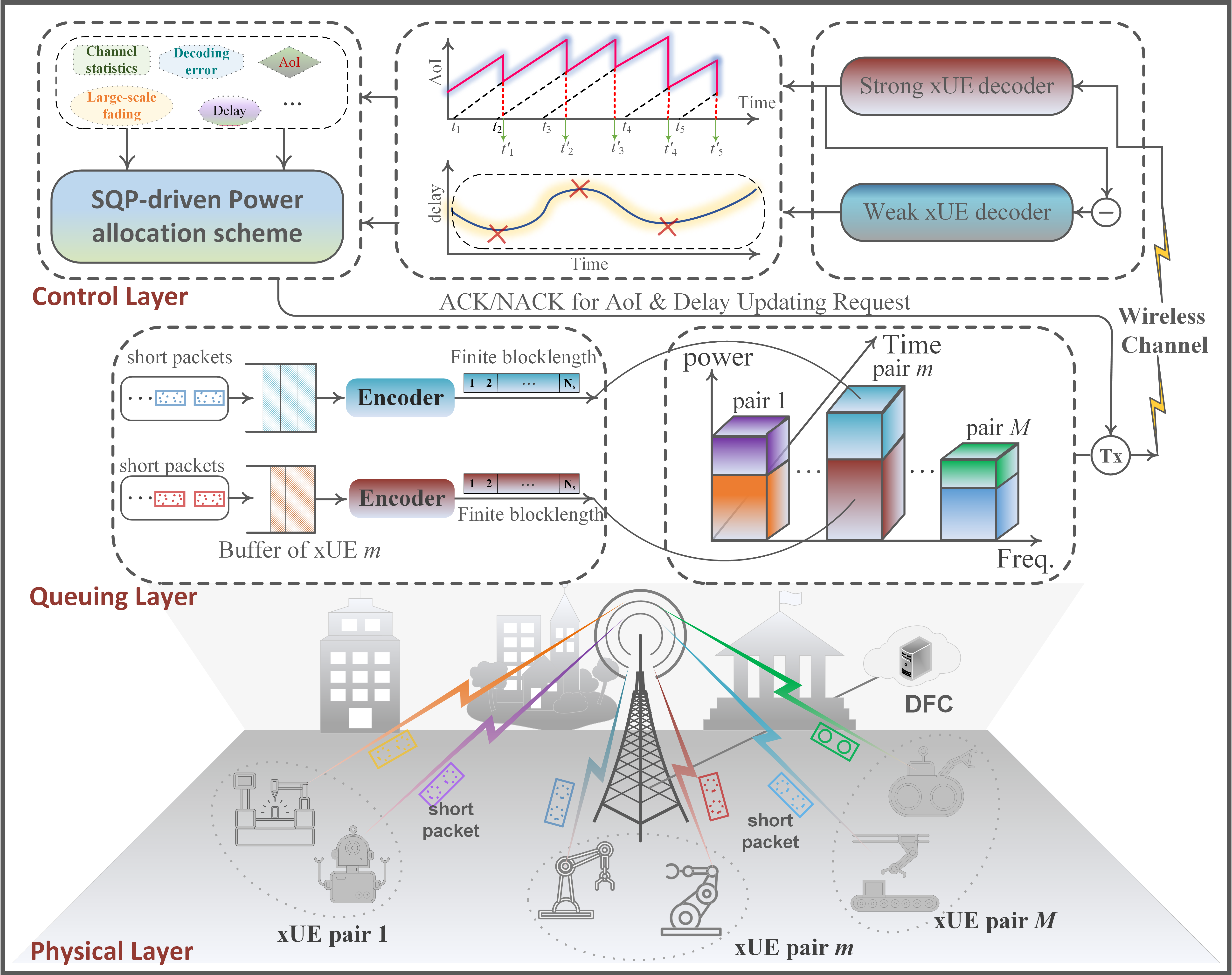}
\vspace{-0.5em}
\caption{The system architecture of the proposed NOMA-assisted uplink xURLLC wireless networks.}
\label{fig:label}
\vspace{-1.5em}
\end{figure*}

\vspace{-0.5em}

\section{NOMA-xURLLC Architecture}

\par As illustrated in Fig. 1, we propose a NOMA-xURLLC network architecture, in which the base station (BS) serves multiple xURLLC user equipments (xUEs) simultaneously by equally dividing the available spectrum resource into several orthogonal subchannels, and all xUEs are uniformly distributed in the cell. The xUEs are sorted and paired in descending order based on their channel qualities, forming one-to-one pairs consisting of a ``weak xUE'' (WU) with a relatively poor channel gain and a ``strong xUE'' (SU). Both xUEs multiplex the same subchannel using NOMA. The proposed NOMA-xURLLC network architecture comprises three main layers: the physical layer, queuing layer, and control layer.

\par \textbf{Physical Layer:} To fulfill the low-latency and high-freshness requirements, xURLLC relies on accommodating vast amounts of short-packet communications in highly time-varying wireless networks\cite{park2022extreme,she2021tutorial,bennis2018ultrareliable}. The size of short packets is typically $100 \sim 200$ bits \cite{she2021tutorial}. In this context, Shannon's capacity is inadequate due to the short timeslots and finite blocklength. In contrast, the finite blocklength coding theory (FBC) with non-vanishing decoding error is preferable\cite{polyanskiy2011feedback0}. In our proposed NOMA-xURLLC network architecture, short packets of each xUE are first encoded into finite blocklengths, then transmitted through wireless fading channels. Particularly, the message space cardinality of each short packet is determined based its size, and a message encoder can map each message into a codeword with a finite length. The coding rate and decoding error probability of each short packet can be obtained based on the FBC theory, respectively\cite{polyanskiy2011feedback0}. The channel coefficient of each xUE includes large-scale fading and small-scale fading, with the BS performing the successive interference cancellation (SIC) technique based on the NOMA principle to decode the messages of the SU and WU. The SINRs of SU and WU can be obtained, with the SU being affected by interference from the WU since its messages are decoded first.

\par \textbf{Queuing Layer:} We consider a SQP-driven short-packet queuing system in the proposed NOMA-xURLLC network architecture. Each xUE maintains a buffer operating on the first-come and first-serve (FCFS) policy, where short packets are removed from the buffer only upon receiving their acknowledgements (ACKs) to ensure ultra-reliability. The real-time monitoring and recording of the arrival, departure, and service processes of short packets are essential for analyzing the tail probability of delay. Specifically, the arrival process represents the short packets put into the buffer within a time slot, while the service process and departure process indicate the number of transmitted short packets and successful arrivals with ACKs in a time slot, respectively. Additionally, the arrival, departure, and service times of short packets are also monitored and recorded to capture the tail probability of AoI. The arrival time describes the packet generation time stamp, the service time reflects the transmission and processing time required for each short packets over the wireless network, and the departure time indicates the time stamp of deleting corresponding short packets from the buffer upon receiving ACKs. Furthermore, the service process and the service time depend on xUEs' channel gains, which vary across different time slots.

\par \textbf{Control Layer:} Each xUE at the sender side senses the feedback messages, such as ACK/NACK and power allocation strategies, and frequently sends newly generated short packets (e.g., task progress, measured data, and device status) to the BS. At the receiver side, the BS also frequently senses and collects messages, which are further sent to the data fusion center (DFC) through the backhaul link for processing. The DFC extracts key data, such as channel statistics, large-scale fading, decoding errors, delay, and AoI, to perform the proposed SQP-driven power allocation to guarantee xURLLC's SQP performance. The power allocation results are fed back to the xUEs through signaling and updated only when large-scale information changes. In this way, the signaling overhead of power allocation for xURLLC can be reduced significantly. In addition, the reliability of xURLLC for each xUE relies on the decoding error information fed back by the DFC, which determines whether to retransmit short packets or not. To this end, the proposed SNC-SQP theoretical framework is deployed in the DFC to estimate the tail probabilities of delay and AoI, which affect the power allocation results of xURLLC and reflect its SQP performance in terms of delay and AoI.

\vspace{-0.6em}

\section{Key Challenges}
\par To achieve a NOMA-xURLLC network architecture with low-latency, ultra-reliability, low-AoI, and energy-efficient, the key challenges and the corresponding potential solutions are elaborated in this section.

\par \textbf{Co-channel Interference:} Although NOMA possesses the potential to bolster the spectrum efficiency and service capacity of xURLLC, its integration inevitably introduces co-channel interference \cite{liu2022optimization,kotaba2020urllc,yin2022power,schiessl2020noma}, which poses a significant impediment to the SQP performance of mission-critical and high-risk xURLLC while concurrently complicating the analysis of the tail distributions for xURLLC's rigorous KPIs. SNC serves as a formidable theoretical methodology for probing the tail distributions of intricate stochastic processes, thereby unlocking the potential to capture reliable tail distributions and provide insightful analysis of SQP performance. However, the prerequisite for harnessing these capacities lies in the initial development of a dependable SNC-based theoretical framework precisely tailored to the context of NOMA-xURLLC with co-channel interference.

\par \textbf{Tail Probability Analysis:} Designing methodologies for 5G URLLC that are interoperable with NOMA-xURLLC network architecture is a challenging task. Although SNC theory holds promise in providing a principled framework for analyzing the xURLLC's KPIs, the statistic of fading channels, the complexity introduced by integrated NOMA, and the stochastic natures of service process and service time in short-packet communications have made it difficult to obtain analytical expressions for the tail distributions within the proposed SNC-based theoretical framework \cite{bennis2018ultrareliable,fidler2014guide,schiessl2020noma}. Additionally, capturing the tail probabilities of delay and AoI has become more challenging. For this reason, it becomes necessary in the developed SNC-based theoretical framework to derive the reliable upper bounds for these tail distribution analytical expressions, converting the unattainable tail distribution into a manageable upper bound, and providing theoretical guidance for the SQP performance analysis of xURLLC.

\vspace{-0.4em}

\par \textbf{Power Allocation:} xURLLC dedicates its efforts to optimizing network performance while upholding the statistical QoS constraints of rigorous KPIs. However, optimization problems involving statistical delay and AoI as constraints typically manifest complex and non-convex structures \cite{yates2021age,zhang2021statistical,she2021tutorial,schiessl2020noma}. Additionally, addressing optimization problems encompassing tail probabilities of delay and AoI derived through SNC-based approaches consistently poses challenges. As a result, thoroughly investigating the power allocation schemes with tail probabilities as constraints can lead to constructive insights for xURLLC. To address these non-convex optimization problems, we can delve into sequential optimization methods, aiming to decompose them into several subproblems that necessitate sequential optimization to achieve equivalent solutions, thereby reducing the complexity associated with solving the power allocation problem.

\vspace{-0.6em}

\section{The SNC-SQP Theoretical Framework}
\par As illustrated in Fig. 2, we design a theoretical framework named SNC-SQP to provide quantitative insights into the tail distributions of xURLLC's KPIs under the impact of co-channel interference in NOMA-xURLLC. The tail distributions captured through SNC-SQP can provide valuable theoretical guidance for the development of power allocation schemes.
\par \textbf{Synthetical statistical QoS requirements (SS-QoS):} We define SS-QoS to quantify the SQP requirements of xURLLC. SS-QoS is composed of target delay, delay violation probability threshold, target AoI, AoI violation probability threshold, and QoS exponent\cite{bennis2018ultrareliable,fidler2014guide,jiang2008stochastic}. In SNC, the QoS exponent is a free parameter greater than zero, which measures the exponential decay rate of the statistical QoS violation probabilities\cite{bennis2018ultrareliable,fidler2014guide,jiang2008stochastic}. NOMA-xURLLC can tolerate arbitrarily loose SQP as the QoS exponent approaches zero, while it cannot tolerate any SQP as the QoS approaches infinity.

\par \textbf{Statistical delay violation probability (SDVP):} SDVP is defined to characterize the tail distribution of delay and evaluate the delay performance of xURLLC. The total delay primarily consists of transmission delay and queuing delay. Specifically, SDVP quantifies the probability that the delay of short-packet communication of xUE exceeds the target delay, subject to the predetermined delay violation probability threshold, for a given set of SS-QoS requirements. This metric provides an important evaluation of reliability of NOMA-xURLLC in meeting latency demands.

\par \textbf{Statistical AoI violation probability (SAVP):} Similarly, SAVP is defined to characterize the tail distribution of AoI and evaluate the freshness performance of short packets. SAVP quantifies the probability that the AoI of xUE’s shortpacket communication exceeds the target AoI, subject to the predetermined AoI violation probability threshold, for a given set of SS-QoS requirements. This metric enables a quantitative evaluation of NOMA-xURLLC’s ability to meet the freshness requirements of real-time applications.

\par Based on SNC theory, obtaining analytical expressions for SDVP and SAVP is typically unattainable due to the stochastic nature of channels and co-channel interference\cite{bennis2018ultrareliable,fidler2014guide,schiessl2020noma}. As a result, we resort to investigating UB-SDVP and UB-SAVP thoroughly. By utilizing SNC theory, we prove that UB-SDVP can be formulated as the infimum of a kernel function concerning the QoS exponent \cite{jiang2008stochastic,fidler2014guide}. This kernel function is uniquely determined by the moment generative function (MGF) of the arrival process and the inverse-MGF of the service process for short packets \cite{jiang2008stochastic,fidler2014guide}. Considering the arrival process follows a Poisson stochastic process, and thus the inter-arrival time of two adjacent short packets follows an exponential process \cite{zhang2021statistical}. Moreover, the service process is influenced by co-channel interference and channel quality, and hence, SNC-SQP requires incorporating the probability density functions (PDFs) of SINRs for both strong and weak xUEs. In Rayleigh fading channels, the PDF of the SINR for a WU can be directly obtained as it experiences no interference. However, for a SU that suffers from interference, the PDF of its SINR is re-derived in SNC-SQP using integral transform methods \cite{schiessl2020noma}.

\par Each short packet has an AoI that is calculated as the sum of its residency time and inter-arrival time, where the residency time is determined by the difference between the packet's departure and arrival times \cite{zhang2021statistical}. Using SNC theory, we prove that UB-SAVP can be uniquely determined by the MGF of the service time and both the MGF and inverse-MGF of the inter-arrival time. To guarantee ultra-reliability of xURLLC, decoding errors caused by channel fluctuations for short packets need to be considered. Based on the automatic repeat request protocol\cite{zhang2021statistical}, the service time can be modeled as the product of the number of channel uses (CUs), the unit time of each CU, and the decoding success probability. The MGF of the service time can be determined by the PDFs of SINRs for both strong and weak xUEs. The inter-arrival time follows an exponential process, and its MGF and inverse-MGF are easily obtained.

\begin{figure}[htbp]
\centering
\includegraphics[scale=0.12]{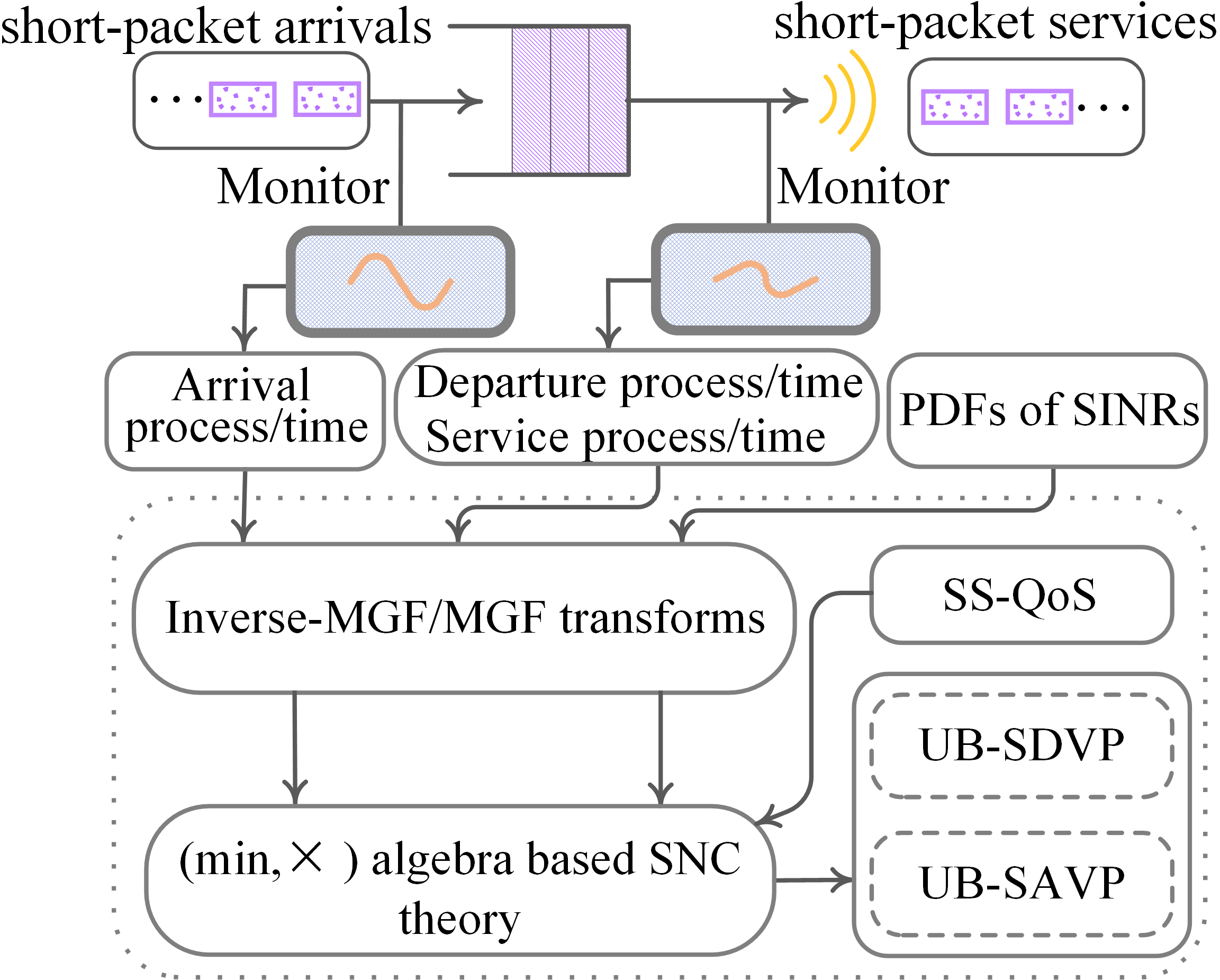}
\vspace{-0.2em}
\caption{The developed SNC-SQP theoretical framework.}
\label{fig:label}
\end{figure}

\vspace{-1.5em}

\section{SQP-Driven Power Allocation and Solutions}

\par The proposed SNC-SQP provides valuable theoretical guidance for developing SQP-driven power allocation in NOMA-xURLLC. We formulate an SQP-driven power optimization problem with the objective of minimizing the uplink transmit power under SS-QoS constraints, subject to four constraints. Firstly, the freshness constraint limits UB-SAVP to a predetermined AoI violation probability threshold, ensuring that short packets remain fresh within the target AoI and guaranteeing reliability of xURLLC's SQP performance in terms of AoI \cite{jiang2008stochastic}. Secondly, the delay constraint limits UB-SDVP to a predetermined delay violation probability threshold, ensuring that short packets are delivered within the target delay and guaranteeing reliability of xURLLC's SQP performance in terms of delay \cite{schiessl2020noma}. The third constraint limits the maximum transmit power to guarantee energy efficiency. The final constraint is the stability condition, which requires that the product of the arrival process's MGF and the service process's inverse-MGF be less than 1 for UB-SDVP, and the product of the service time's MGF and the inter-arrival time's inverse-MGF to be less than 1 for UB-SAVP \cite{fidler2014guide,jiang2008stochastic,schiessl2020noma}.

\par To effectively tackle this intractable non-convex problem, we deeply explore its intrinsic properties from the perspective of its structure. The intrinsic properties of this problem are summarized as follows:
\begin{itemize}
  \item \textbf{\emph{Claim 1:}} \emph{UB-SDVP and UB-SAVP, as well as their stability conditions are convex functions with respect to QoS exponent.}
  \item \textbf{\emph{Claim 2:}} \emph{For a WU, both UB-SDVP and UB-SAVP are monotonically decreasing functions with respect to its own transmit power, and are independent of the SU's transmit power. For a SU, both UB-SDVP and UB-SAVP are also monotonically decreasing functions with respect to its own transmit power, but monotonically increasing with respect to the WU's transmit power.}
  \item \textbf{\emph{Claim 3:}} \emph{Both the inverse-MGF of service process and the MGF of service time are convex functions with respect to transmit power.}
\end{itemize}

\par Based on the above three claims, the original problem can be equivalently decomposed into two subproblems. Subproblem I involves optimizing the uplink transmit power for WU and is a convex problem, while subproblem II involves optimizing the uplink transmit power for SU and is non-convex. Since WU experiences no interference, while SU suffers from co-channel interference, the two subproblems must be solved in sequence; in other words, subproblem I must be solved first, followed by subproblem II. According to \emph{Claim 1}, given the transmit powers of WU and SU, the feasible regions of QoS exponent for both WU and SU can be determined using the Golden-Section search method, and then the QoS exponents that minimize both UB-SDVP and UB-SAVP within the feasible regions can be determined using the Gradient-Descent method. On this basis, subproblem I can be efficiently solved, and the optimal uplink transmit power for WU can be obtained by using Bisection search method. Based on \emph{Claim 2} and \emph{Claim 3}, the co-channel interference experienced by the SU is minimized when the transmit power of WU is at its minimum, and subproblem II degenerates into a convex problem. The same solving procedure can be applied to obtain SU's optimal uplink transmit power.

\vspace{-0.8em}
\section{Performance Evaluation}
\par In this section, extensive simulations are conducted to validate the proposed NOMA-xURLLC network architecture. The simulation parameters include a cell service radius of $500$ meters, a short packet size (PS) of $100$$\sim$$200$ bits, a bandwidth of $2$ MHz per orthogonal subcarrier, a time slot of $1$ ms, an average arrival rate of $200$$\sim$$300$ kbps, a maximum uplink transmit power of $2.0$ W for each xUE, and a noise power of $-176$ dBm/Hz. The small-scale fading of channels follows a Rayleigh fading distribution. WU is interference-free, while SU experiences co-channel interference from WU. The path loss exponent is 2.5, and the shadow fading is modeled as a lognormal distribution with a standard variance of $8$ dB. The effectiveness of the proposed NOMA-xURLLC are evaluated from two perspectives. Firstly, the Monte-Carlo method is utilized to generate $10^{8}$ random channels to estimate the actual SDVP and SAVP, which are compared with the derived UB-SDVP and UB-SAVP to validate the accuracy and reliability of the designed SNC-SQP. Secondly, we demonstrate the advantages of the optimal power allocation scheme by comparing it with conventional schemes in OMA.

\begin{figure}[htbp]
\centering
  \subfigure[]{
    \includegraphics[scale=0.043]{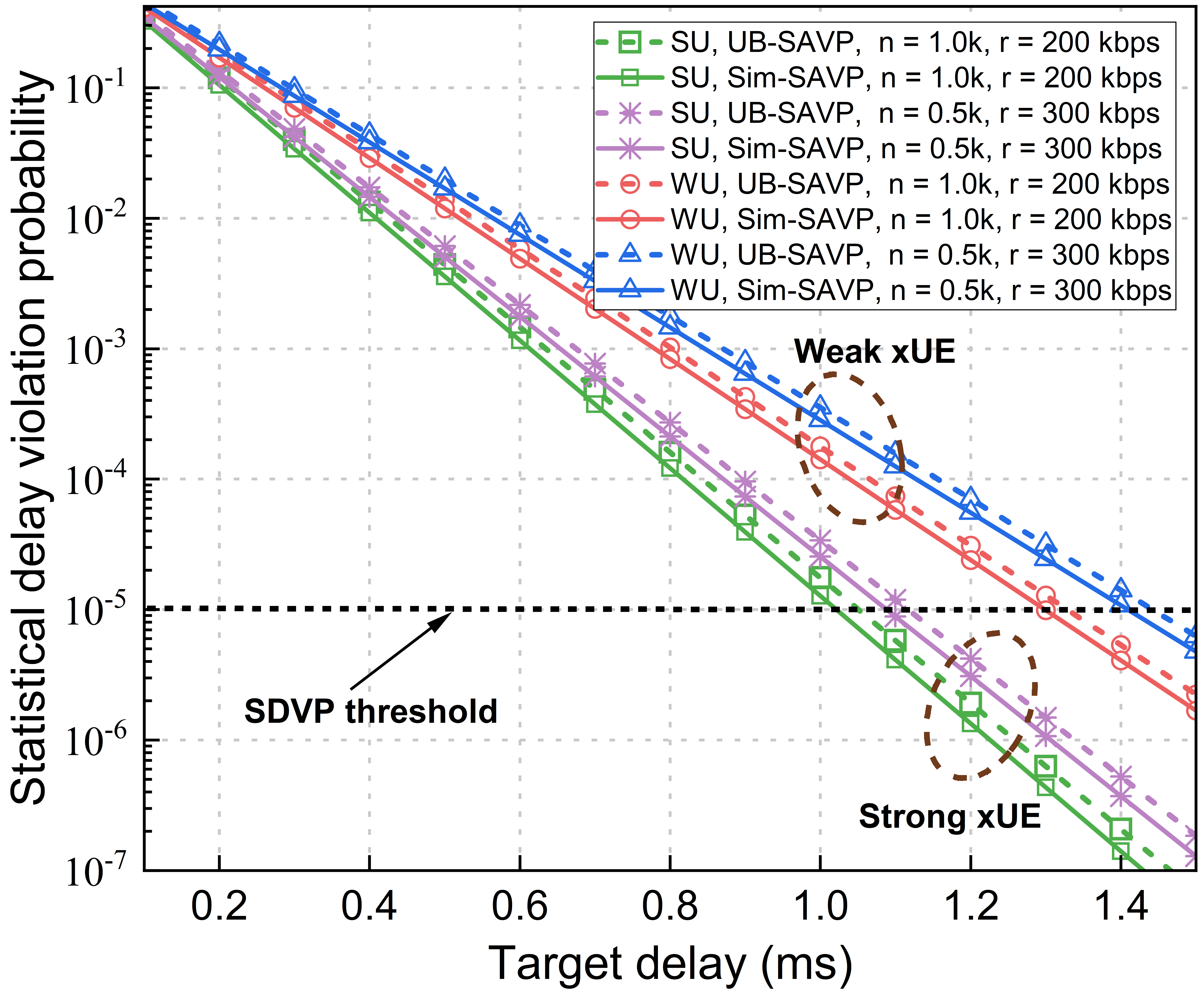}
      \vspace{-1em}
  }
  \subfigure[]{
    \includegraphics[scale=0.030]{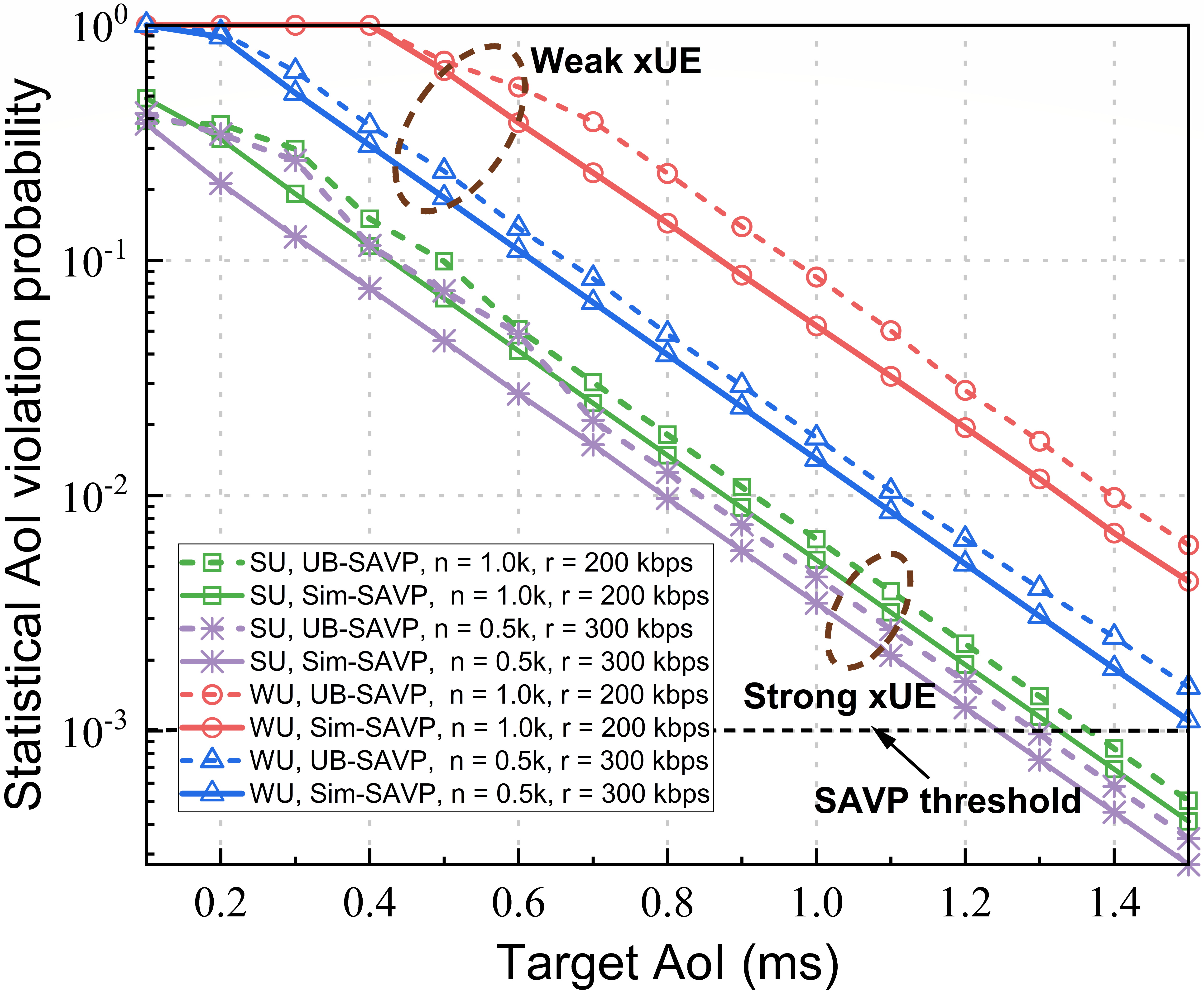}
  }
  \vspace{-0.3em}
\caption{Validation of the accuracy and reliability of the SNC-SQP theoretical framework.}
\label{fig:label}
\vspace{-1.0em}
\end{figure}

\par Fig. 3 demonstrates the accuracy and reliability of the proposed SNC-SQP by comparing the UB-SDVP and UB-SAVP with the Sim-SDVP and Sim-SAVP estimated using the Monte-Carlo method. We set the packet size (PS) to $150$ bits, the SDVP threshold to $10^{-5}$, and the SAVP threshold to $10^{-3}$. The results reveal that the SNC-SQP is effective in analyzing the tail distributions of xURLLC's KPIs, as demonstrated by the similarity in the slopes of the curves produced by SNC-SQP and the Monte-Carlo method under various conditions. Furthermore, a horizontal gap exists between upper bounds and simulations for SDVP and SAVP, which is due to the SNC-SQP being based on $(min,\times)$-SNC theory that captures the queuing system's nonasymptotic statistical performance bounds. Therefore, it converts the unattainable SDVP and SAVP into manageable upper bounds, providing theoretical guidance for SQP-driven power optimization. We observe that the arrival rate is crucial in xURLLC, as it impacts SDVP performance even with a shorter blocklength, resulting in reduced performance under limited resources. This is because an increase in arrival rate leads to a longer average service time required for short packets. Additionally, moderately increasing the arrival rate can enhance SAVP performance because of the tradeoff between the arrival rate and the AoI \cite{park2022extreme,yates2021age,zhang2021statistical}, making it beneficial for maintaining the freshness of short packets.

\vspace{-0.2em}

\begin{figure}[h]
\centering
\includegraphics[scale=0.030]{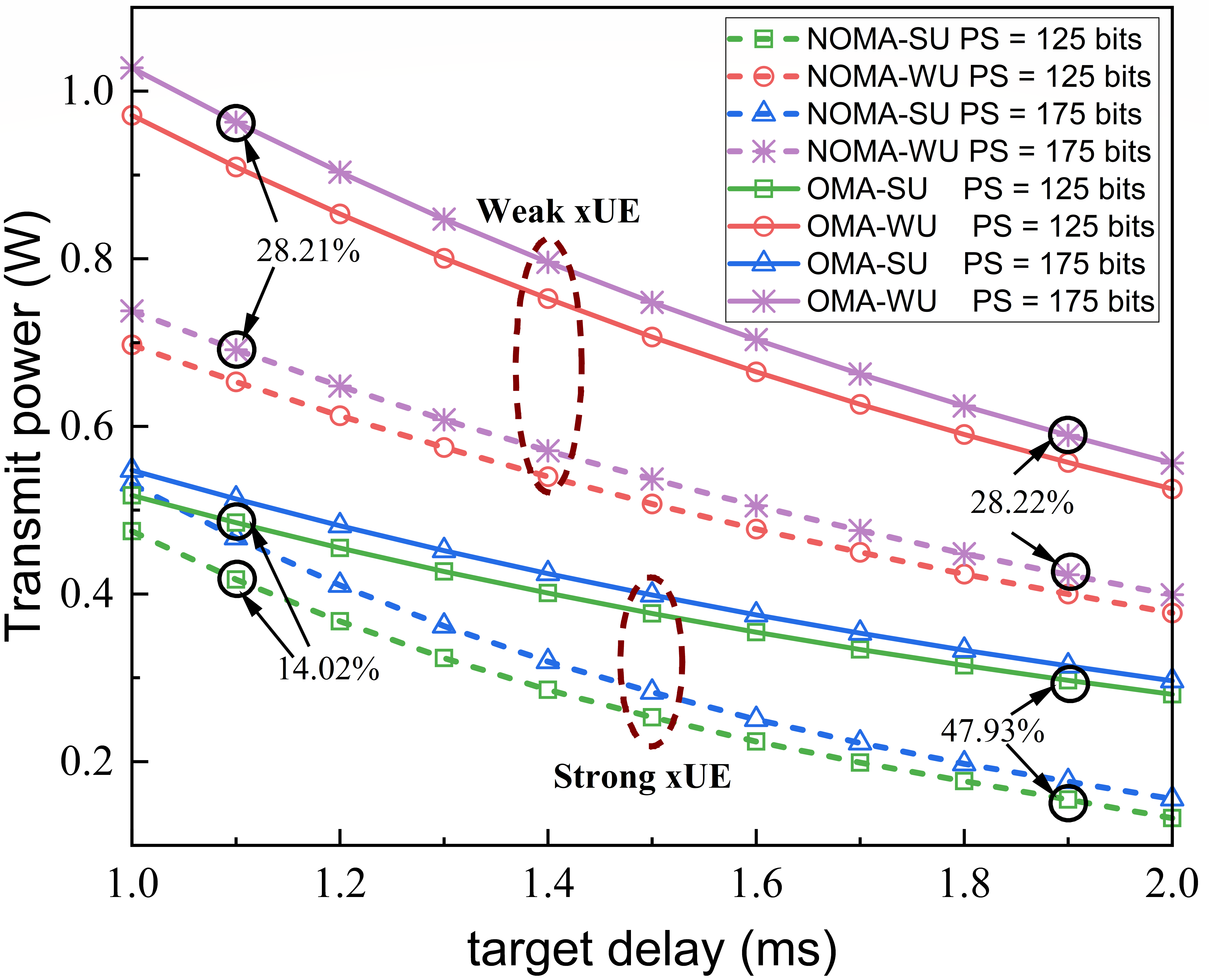}
\vspace{-0.5em}
\caption{Optimal power allocation versus target delay.}
\label{fig:label}
\vspace{-1.0em}
\end{figure}

\par In Fig. 4, we investigate the optimal power allocation for different target delays in the finite blocklength regime. The average arrival rate, SDVP threshold, and SAVP threshold are set to $250$ kbps, $10^{-5}$, and $10^{-3}$, respectively. Numerical results show that the proposed scheme consistently consumes less transmit power than the OMA scheme for various target delays, indicating its superiority. For a WU, the proposed scheme achieves similar performance gains compared to the OMA scheme across different target delays, with a performance improvement of around 28.21 $\%$ for a PS of $175$ bits and a target delay of $1.1$ ms, and about 28.22$\%$ for a target delay of $1.9$ ms. This is because WUs experience no interference whether NOMA or OMA are used, and the performance improvement stems from NOMA's advantage in high-frequency spectrum utilization. For a SU, our scheme's performance improvement compared to the OMA scheme is quite limited when the target delay is stringent, but it achieves substantial gains when the target delay is relatively loose. For example, with a PS of $125$ bits and a target delay of $1.1$ ms, the proposed scheme reduces power consumption by approximately 14.02$\%$ compared to the OMA, while at a target delay of $1.9$ ms, it achieves about 47.93$\%$ improvement. This is due to SUs' susceptibility to co-channel interference, which can significantly degrade the SDVP performance when the target delay is stringent. Therefore, increasing the transmit power is necessary to guarantee reliability. However, when the target delay is relatively loose, NOMA can significantly improve resource utilization and save transmit power while maintaining SDVP performance for SUs.

\par In Fig. 5, we investigate the optimal power allocation for different target delays in the finite blocklength regime. Numerical results show that both the proposed and OMA schemes tend to allocate the maximum transmission power when the target AoI is relatively stringent (e.g., 1.0$\sim$1.2 ms) to guarantee the performance of SDVP and SAVP, indicating the importance of AoI in xURLLC's resource allocation. Compared to delay, PS has a greater impact on AoI, and the proposed scheme consistently outperforms OMA when PS decreases from $175$ bits to $125$ bits. Similar to Fig. 4, the proposed SQP-driven power allocation scheme yields better power performance under NOMA than OMA for WUs. Interestingly, for a SU under stringent target AoI, NOMA consumes more transmit power than OMA. For instance, with a PS of $125$ bits and a target AoI of $1.4$ ms, NOMA consumes 17.15$\%$ more transmit power than OMA. As the target AoI becomes looser, NOMA's advantages gradually emerge. Notably, with a target AoI of 1.6 ms as the turning point, the performance of NOMA surpasses OMA. For a target AoI of $1.8$ ms, the performance improvement of NOMA compared to OMA exceeds 49.85$\%$. This is mainly due to xURLLC's sensitive to interference under stringent target AoIs, which impedes energy-efficient and high-freshness short-packet communications. Moreover, the proposed SQP-driven power allocation scheme overall outperforms the OMA scheme by a large margin.

\vspace{-0.2em}

\begin{figure}[h]
\centering
\includegraphics[scale=0.030]{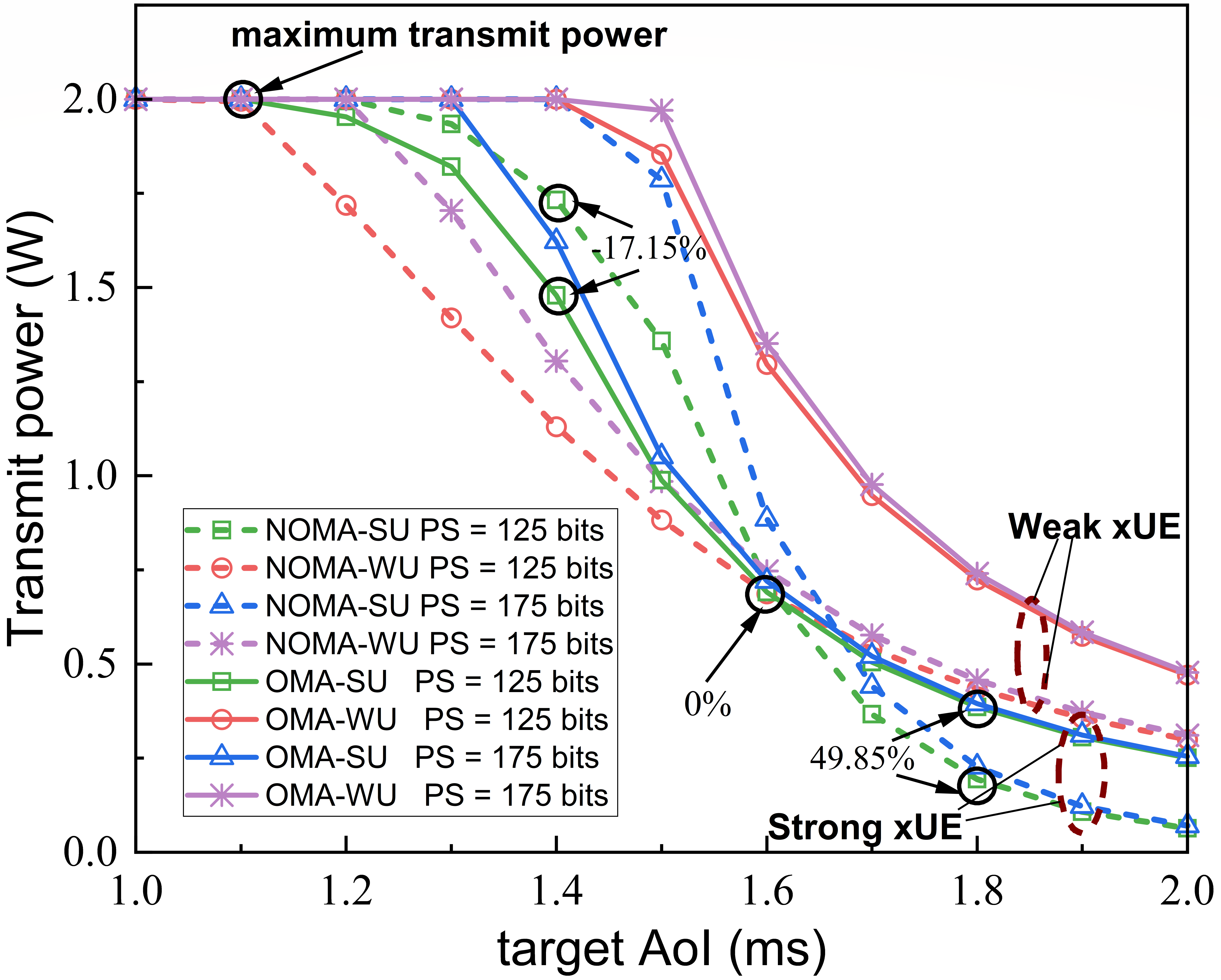}
\vspace{-0.5em}
\caption{Optimal power allocation versus target AoI.}
\label{fig:label}
\vspace{-1.0em}
\end{figure}

\vspace{-0.5em}

\section{Conclusion and Future Work}

\par In this article, we have proposed an innovative NOMA-xURLLC network architecture that integrates NOMA transmission methods into xURLLC for efficient short-packet communications. Leveraging the unique features of xURLLC, we have introduced SNC theory to design an SNC-SQP theoretical framework that accurately captures the tail distributions of xURLLC's KPIs. Specifically, we have defined UB-SDVP and UB-SAVP to characterize the reliability of xURLLC in terms of delay and AoI. Based on the designed theoretical framework, we have proposed an SQP-driven power optimization problem that aims to minimize the transmit power while guaranteeing ultra-reliable, low-latency, high-freshness, and energy-efficient short-packet communications. Extensive simulations have demonstrated the accuracy of the designed theoretical framework in capturing the tail distributions of xURLLC's KPIs, and have verified that the proposed power allocation scheme can minimize the xURLLC's uplink transmit power while ensuring its SQP performance. Additionally, compared to conventional OMA schemes, the proposed optimal power allocation scheme has significantly improved the uplink transmit power and has guaranteed better SQP performance for xURLLC.

\vspace{-0.3em}

\par As an emergent network architecture, there are many open issues for further research in the future:

\noindent \textbf{Flexible Interference Management for Massive xURLLC:} Essentially, NOMA epitomizes an extreme interference management paradigm, where the former treats interference as noise while the latter undertakes complete interference \cite{dai2015non,10038476}. As an extension of NOMA, Rate-splitting multiple-access (RSMA) empowers multiple xURLLC devices with concurrent non-orthogonal interference management potential by partially treating interference as noise while partially decoding it \cite{10038476}. This endows xURLLC devices with heightened flexibility in navigating intricate interference scenarios. Consequently, RSMA holds promise in achieving harmonious equilibriums among latency, reliability, and scalability for massive xURLLC deployment \cite{she2021tutorial}.

\noindent \textbf{Security of xURLLC:} In xURLLC scenarios, the indispensability of short-packet communications is underscored by the stringent QoS demands for ultra-reliable, low-latency, high-freshness, and energy efficiency. However, the prevalence of complex cryptographic algorithms hinders their suitability for computationally limited xURLLC devices. Consequently, in the finite blocklength regime, the vulnerability to attack and theft is significantly exacerbated. To this end, the dedicated design of a physical layer security mechanism tailored for xURLLC becomes exceedingly valuable in ensuring the integrity, privacy, and stability of data processing by enabling maximum secure communication rate over eavesdropping channels in the short blocklength regime.

\noindent \textbf{Predictive xURLLC and Machine Learning:} In 6G networks, revolutionary applications like metaverse exemplify a seamless integration of ultra-high data rates, and ultra-reliable and low-latency haptic feedbacks \cite{bennis2018ultrareliable}. To accomplish this, precise measurement and prediction of haptic feedbacks are imperative. Leveraging machine learning (ML) for delay estimation and AoI measurement holds immense potential for enhancing the reliability of xURLLC. However, due to the complexity of xURLLC's samples, training ML prediction models becomes challenging. In this context, federated learning (FL) emerges as a viable solution by facilitating periodic exchange of locally trained model parameters, thereby mitigating the prediction costs for xURLLC \cite{she2021tutorial}. Additionally, to assess the reliability of predictions for unseen training samples, quantifying the generalization error of xURLLC's samples becomes an essential endeavor \cite{park2022extreme}.

\vspace{-0.8em}
\section*{Acknowledgments}
\par This work was supported by the National Science Foundation of China (NSFC) (No. U21A20452, No. U19B2044).

\vspace{-0.7em}
\footnotesize
\bibliographystyle{IEEEtran}
\bibliography{IEEEabrv,ref}
\vspace{-1.5em}
\section*{Biographies}
\noindent
\footnotesize{{\sc Yuang Chen} is currently pursuing the Ph.D. degree in the Department of EEIS, USTC. His research interests include NOMA, RSMA, and xURLLC.\\

\vspace{-0.2em}

\noindent
\footnotesize{{\sc Hancheng Lu} is currently a Professor with the Department of EEIS, USTC. His research interests include resource optimization in wireless communication systems, such as URLLC, SNC, RIS, NOMA, cell-free, UCN, and caching and service offloading at wireless network edges.}\\

\vspace{-0.22em}

\noindent
\footnotesize{{\sc Langtian Qin} is currently pursuing the master’s degree with the Department of EEIS, USTC, Hefei, China. His research interests include mobile MEC, and UCN.}\\

\vspace{-0.22em}

\noindent
\footnotesize{{\sc Yansha Deng} is currently a Senior Lecturer (Associate Professor) with the Department of Engineering in King's College London. Her research interests include machine learning for 5G/6G wireless networks and molecular communications.}\\

\vspace{-0.22em}

\noindent
\footnotesize{{\sc Arumugam Nallanathan} is Professor of Wireless Communications and Head of the Communication Systems Research (CSR) group in the School of Electronic Engineering and Computer Science at Queen Mary University of London. His research interests include Beyond 5G Wireless Networks, Internet of Things, and Molecular Communications. He is an IEEE Fellow and IEEE Distinguished Lecturer.}\\

\vspace{-0.22em}

\end{document}